\documentclass[prl,twocolumn,showpacs,superscriptaddress,preprintnumbers,floatfix]{revtex4-1}

\usepackage{ifpdf}
\usepackage{hyperref}
\usepackage{dcolumn}
\usepackage{url}
\usepackage{amsmath}
\usepackage{amssymb}
\usepackage{bm}   
\usepackage{bbm}   
\usepackage{verbatim}
\usepackage{stmaryrd}
\usepackage{amsthm}
\usepackage{xcolor}

\ifpdf
  \usepackage[pdftex]{graphicx}         
\else
  \usepackage[dvips]{graphicx}
\fi

\theoremstyle{plain}    \newtheorem{Lem}{Lemma}
\theoremstyle{plain}    \newtheorem*{ProLem}{Proof}
\theoremstyle{plain} 	\newtheorem{Cor}{Corollary}
\theoremstyle{plain} 	\newtheorem*{ProCor}{Proof}
\theoremstyle{plain} 	\newtheorem{The}{Theorem}
\theoremstyle{plain} 	\newtheorem*{ProThe}{Proof}
\theoremstyle{plain} 	\newtheorem{Prop}{Proposition}
\theoremstyle{plain} 	\newtheorem*{ProProp}{Proof}
\theoremstyle{plain} 	
\theoremstyle{plain}	\newtheorem*{Rem}{Remark}
\theoremstyle{plain}	\newtheorem*{Def}{Definition} 
\theoremstyle{plain}	

\newcommand{\eM}     {$\epsilon$\protect\nobreakdash-machine}
\newcommand{\eMs}    {$\epsilon$\protect\nobreakdash-machines}
\newcommand{\EM}     {$\epsilon$\protect\nobreakdash-Machine}

\newcommand{\MeasAlphabet}	{\mathcal{A}}
\newcommand{\MeasSymbol}   { {X} }
\newcommand{\meassymbol}   { {x} }
\newcommand{\BiInfinity}	{ \overleftrightarrow {\MeasSymbol} }

\newcommand{\Past}	{ \overleftarrow {\MeasSymbol} }
\newcommand{\past}	{ {\overleftarrow {\meassymbol}} }

\newcommand{\Future}	{ \overrightarrow{\MeasSymbol} }
\newcommand{\future}	{ \overrightarrow{\meassymbol} }

\newcommand{\AllPasts}	{ { \overleftarrow {\rm {\bf \MeasSymbol}} } }

\newcommand{\CausalState}	{ \mathcal{S} }

\newcommand{\causalstate}	{ \sigma }
\newcommand{\CausalStateSet}	{ \boldsymbol{\CausalState} }
\newcommand{\AlternateState}	{ {\cal R} }

\newcommand{\PrescientState}	{ \widehat{\AlternateState} }

\newcommand{\PrescientStateSet}	{ \boldsymbol{\PrescientState}}

\newcommand{\Prob}		{ {\Pr}}

\newcommand{\Cmu}		{ {C_\mu}}

\newcommand{\EE}		{ {\bf E}}

\newcommand{\FinPast}[1]	{ \overleftarrow {\MeasSymbol}^{#1} }
\newcommand{\finpast}[1]  	{ \overleftarrow {\meassymbol}^{#1} }
\newcommand{\finpastp}[1]  	{ \overleftarrow {\meassymbol}^{\prime{#1}} }
\newcommand{\FinFuture}[1]	{ \overrightarrow{\MeasSymbol}^{#1} }
\newcommand{\finfuture}[1]	{ \overrightarrow{\meassymbol}^{#1} }
\newcommand{\finfuturep}[1]	{ \overrightarrow{\meassymbol}^{\prime{#1}} }

\newcommand{\FutureCausalState}	{ {\CausalState}^+ }

\newcommand{\PastCausalState}	{ {\CausalState}^- }

\newcommand{\FutureCausalStateSet}	{ {\CausalStateSet}^+ }
\newcommand{\PastCausalStateSet}	{ {\CausalStateSet}^- }

\newcommand{\eMachine}	{ M }
\newcommand{\FutureEM}	{ {\eMachine}^+ }
\newcommand{\PastEM}	{ {\eMachine}^- }

\newcommand{\Futurehmu}	{ h_\mu^+ }
\newcommand{\Pasthmu}	{ h_\mu^- }
\newcommand{\FutureCmu}	{ C_\mu^+ }
\newcommand{\PastCmu}	{ C_\mu^- }

\newcommand{\FutureEps}	{ \epsilon^+ }
\newcommand{\PastEps}	{ \epsilon^- }

\newcommand{\PastSim}	{ \sim^- }

\newcommand{\CSjoint}[1][,]{
   \edef\tempa{:}
   \edef\tempb{#1}
   \ifx\tempa\tempb
      \ensuremath{\FutureCausalState\!#1\PastCausalState}
   \else
      \ensuremath{\FutureCausalState#1\PastCausalState}
   \fi
}

\newcommand{\CSjointKL}[3][,]{
   \edef\tempa{:}
   \edef\tempb{#1}
   \ifx\tempa\tempb
      \ensuremath{\FutureCausalState_{#2}\!#1\PastCausalState_{#3}}
   \else
      \ensuremath{\FutureCausalState_{#2}#1\PastCausalState_{#3}}
   \fi
}

\def\clap#1{\hbox to 0pt{\hss#1\hss}}

\def\mathclap{\mathpalette\mathclapinternal}

\def\mathclapinternal#1#2{%
\clap{$\mathsurround=0pt#1{#2}$}}

\begin{document}

\title{The Past and the Future in the Present}

\author{James P. Crutchfield}
\email{chaos@ucdavis.edu}
\affiliation{Complexity Sciences Center and Physics Department,
University of California at Davis, One Shields Avenue, Davis, CA 95616}
\affiliation{Santa Fe Institute, 1399 Hyde Park Road, Santa Fe, NM 87501}

\author{Christopher J. Ellison}
\email{cellison@cse.ucdavis.edu}
\affiliation{Complexity Sciences Center and Physics Department,
University of California at Davis, One Shields Avenue, Davis, CA 95616}

\date{\today}

\bibliographystyle{unsrt}

\begin{abstract}
We show how the shared information between the past and future---the excess
entropy---derives from the components of directional information stored in
the present---the predictive and retrodictive causal states. A detailed proof
allows us to highlight a number of the subtle problems in estimation and
analysis that impede accurate calculation of the excess entropy.
\end{abstract}

\preprint{Santa Fe Institute Working Paper 10-11-XXX}
\preprint{arxiv.org:1011.XXXX [cond-mat.stat-mech]}

\maketitle




\section{Introduction}

Predicting and modeling a system are distinct, but intimately related goals.
Leveraging past observations, prediction attempts to make correct statements
about what the future will bring, whereas modeling attempts to express the
mechanisms behind the observations. In this view, building a model from
observations is tantamount to decrypting a system's hidden organization. The
cryptographic view rests on the result that the apparent information shared between past and future---the excess entropy, which sets the bar for
prediction---is only a function of the hidden stored information---the
statistical complexity \cite{Crut08a}.

The excess entropy, and related mutual information quantities, though, are
widely used diagnostics for complex systems, having been applied to detect
the presence of organization in dynamical 
systems~\cite{Fras86a,Casd91a,Spro03a,Kant06a}, in
spin systems~ \cite{Crut97a,Erb04a}, in neurobiological
systems~\cite{Tono94a,Bial00a}, and even in human language~\cite{Ebel94c,Debo08a}.

For the first time, Ref.~\cite{Crut08a} connected the observed sequence-based
measure, the excess entropy, to a system's internal structure and information
processing. One consequence of the connection, and so our ability to
differentiate between them, is that the excess entropy is an inadequate measure
of a process's organization. One must build models.

Our intention here is rather prosaic, however. We provide a focused and detailed
proof of this relationship, which appears as Thm.~1 in Ref.~\cite{Crut08a} in a
necessarily abbreviated form. A proof also appears in Ref.~\cite{Crut08b}
employing a set of manipulations, developed but not laid out explicitly there,
that require some facility with four-variable mutual informations and with
subtle limiting properties of stochastic processes. The result is that directly
expanding either of these concise proofs, without first deriving the rules,
leads to apparent ambiguities.

The goal in the following is to present a step-by-step proof, motivating and
explaining each step and attendant difficulties. The development also allows
us to emphasize several new results that clarify the challenges in analytically
calculating and empirically estimating these quantities. To get started,
we give a minimal summary of the required background, assuming familiarity
with Refs.~\cite{Crut08a} and \cite{Crut08b}, information theory \cite{Cove06a},
and information measures \cite{Yeun91a}.

\section{Background}

A process $\Prob(\Past,\Future)$ is a \emph{communication channel} with
a fixed input distribution $\Prob(\Past)$:
It transmits information from the \emph{past}
$\Past = \ldots \MeasSymbol_{-3} \MeasSymbol_{-2} \MeasSymbol_{-1}$ to the
\emph{future} $\Future = \MeasSymbol_0 \MeasSymbol_1 \MeasSymbol_2 \ldots$
by storing it in the present. $\MeasSymbol_t$ denotes the discrete random
variable at time $t$ taking on values from an alphabet $\MeasAlphabet$. A
prediction of the process is specified by a distribution $\Prob(\Future|\past)$ 
of possible futures $\Future$ given a particular past $\past$. At a minimum,
a good predictor---call it $\PrescientState$---must capture \emph{all} of a
process's \emph{excess entropy}~\cite{Crut01a}---the information $I$ shared
between past and future: $\EE = I[\Past;\Future]$. That is, for a
good predictor: $\EE = I[\PrescientState;\Future]$.

Building a model of a process is more demanding than developing a prediction
scheme, though, as one wishes to express a process's mechanisms and internal
organization. To do this, computational mechanics introduced an equivalence
relation $\past \sim \past^\prime$ that groups all histories which give rise
to the same prediction. The result is a map
$\epsilon: \AllPasts \to \CausalStateSet$ from pasts to \emph{causal states}
defined by:
\begin{align}
\epsilon(\past) =
  \{ \past^\prime: \Prob(\Future|\past) = \Prob(\Future|\past^\prime) \} ~.
\label{eq:CausalEquiv}
\end{align}
In other words, a process's causal states are equivalence
classes---$\CausalStateSet = \Prob(\Past,\Future) / \sim$---that partition the
space $\AllPasts$ of pasts into sets which are predictively equivalent. The
resulting model, consisting of the causal states and transitions, is called
the process's \emph{\eM} \cite{CompMechMerge}.
Out of all optimally predictive models $\PrescientStateSet$ resulting from a partition of the past, 
the \eM\ captures the minimal amount of information that a process must
store---the \emph{statistical complexity} $\Cmu \equiv H[\CausalState]$.

Said simply, $\EE$ is the effective information transmission rate of the process,
viewed as a channel, and $\Cmu$ is the sophistication of that channel. In
general, the explicitly observed information $\EE$ is only a lower bound on
the information $\Cmu$ that a process stores \cite{CompMechMerge}.

The original development of \eMs\ concerned using the past to predict the
future. One can, of course, use the future to retrodict the past by scanning
the measurement variables in the reverse-time direction, as opposed to the
default forward-time direction. With this in mind, the original map $\epsilon(\cdot)$
from pasts to causal states is denoted $\FutureEps$ and it gave, what are
called, the \emph{predictive} causal states $\FutureCausalStateSet$. When
scanning in the reverse direction, we have a new equivalence relation,
$\future \PastSim \future^\prime$, that groups futures which are equivalent for
the purpose of retrodicting the past:
$\PastEps(\future) =
  \{ \future^\prime: \Prob(\Past|\future) = \Prob(\Past|\future^\prime) \}$.
It gives the \emph{retrodictive} causal states
$\PastCausalStateSet = \Prob(\Past,\Future) / \PastSim$.

In this bidirectional setting we have the forward-scan \eM\ $\FutureEM$ and its
reverse-scan \eM\ $\PastEM$. From them we can calculate corresponding entropy
rates, $\Futurehmu$ and $\Pasthmu$, and statistical complexities,
$\FutureCmu \equiv H[\FutureCausalState]$ and
$\PastCmu \equiv H[\PastCausalState]$, respectively. Notably, while a stationary
process is equally predictable in both directions of
time---$\Futurehmu = \Pasthmu$---the amount of stored information
differs in general: $\FutureCmu \neq \PastCmu$ \cite{Crut08a}.

Recall that Thm.~1 of Ref.~\cite{Crut08a} showed that the shared
information between the past $\Past$ and future $\Future$ is the mutual
information between the predictive ($\FutureEM$'s) and
retrodictive ($\PastEM$'s) causal states:
\begin{equation}
\EE = I[\CSjoint[;]] ~.
\end{equation}
This led to the view that the process's \emph{channel utilization}
$I[\Past;\Future]$ is the same as that in the channel between a process's
forward and reverse causal states.

To understand how the states of the forward and reverse \eMs\ capture
information from the past and the future---and to avoid the
ambiguities alluded to earlier---we must analyze a four-variable
mutual information:
$I[\Past;\Future;\CSjoint[;]]$.
A large number of expansions of this quantity are possible. A systematic
development follows from Ref.~\cite{Yeun91a} which showed that Shannon entropy
$H[\cdot]$ and mutual information $I[\cdot\,;\cdot]$ form a signed measure 
over the space of events.

\section{Two Issues}

The theorem's proof can be expressed in a very compact way using
several (implied) rules:
\begin{align}
\EE & = I[\Past;\Future]
	\label{eq:ShortProofStep1} \\
  & = I[\FutureEps(\Past); \PastEps(\Future) ]
	\label{eq:ShortProofStep2} \\
  & = I[\FutureCausalState; \PastCausalState ] ~.
\end{align}
While this proof conveys the essential meaning and, being short, is easily
intuited, there are two issues with it. The concern is that, if the concise
proof is misinterpreted or the rules not heeded, confusion arises. Refs.
\cite{Crut08a} and \cite{Crut08b} develop the appropriate rules, but do not
lay them out explicitly.

The first issue is that naive expansion of the past-future mutual informations leads
to ambiguously interpretable quantities. The second issue is that implicitly
there are Shannon entropies---e.g., $H[\Past]$ and $H[\Future]$---over
semi-infinite chains of random variables and these entropies diverge in the
general case. Here, via an exegesis of the concise proof, we show how to
address these two problems and, along the way, explicate several of the required
rules. We diagnose the first issue and then provide a new step-by-step proof,
ignoring the second issue of divergent quantities. We end by showing how to
work systematically with divergent entropies.

\section{Comparable Objects and Sufficiency}

The first problem comes from inappropriate application of the
$\epsilon(\cdot)$ functions. The result is the inadvertent introduction of
incomparable quantities. Namely,
\begin{align}
\Prob(\Future|\Past) & = \Prob(\Future|\FutureEps(\Past)) \nonumber \\
  & = \Prob(\Future|\FutureCausalState)
\label{eq:Sufficient}
\end{align}
is a proper use of the predictive causal equivalence relation---the
probabilities at each stage refer to the same object, the future $\Future$.
We say that the predictive causal states are sufficient statistics
for the future.

The following use (in the first equality) is incorrect, however:
\begin{align*}
\Prob(\Past) & = \Prob(\FutureEps(\Past)) \\
  & = \Prob(\FutureCausalState) ~,
\end{align*}
even though it appears as a straightforward (and analogous) application of the
causal equivalence relation. The problem occurs since the first equality
incorrectly conflates probability of two different objects---a future and a
causal state; an element and a set. A handy mnemonic for the appearance of this error is to interpret
the expression literally: typically, a causal state has positive probability,
but an infinite future has zero probability. Clearly, a wrong statement.

There are restrictions on when the causal equivalence relation can be
applied. In particular, in the shorthand proof of Thm.~\ref{EasCausalMI}
above, there are ambiguous expansions of the mutual information that lead one
to such errors. These must be avoided.

Specifically, the step~(Eq.~(\ref{eq:ShortProofStep2})) involving the
simultaneous application of the forward
and reverse causal equivalence relations must be done with care. Here, we show
how to do this. But, first, let's explore the problem a bit more. Starting
from Eq.~(\ref{eq:ShortProofStep1}), we go one step at a time:
\begin{align}
I[\Past;\Future] & = I[\FutureEps(\Past);\Future] \nonumber \\
  & = I[\FutureCausalState;\Future] ~.
\label{eq:FirstApplicationOfEps}
\end{align}
The result is correct, largely because one has in mind the more detailed series
of steps using the mutual information's component entropies. That is, let's
redo the preceding:
\begin{align}
I[\Past;\Future] & = H[\Future] - H[\Future|\Past] \nonumber \\
  & = H[\Future] - H[\Future|\FutureEps(\Past)]
  \label{eq:SecondApplicationOfEps} \\
  & = H[\Future] - H[\Future|\FutureCausalState] \nonumber \\
  & = I[\FutureCausalState;\Future]  ~.
\label{eq:FStateFuture}
\end{align}
Notice that the application of $\FutureEps(\cdot)$ occurs only in conditioning.
Also, for the sake of argument, we temporarily ignore the appearance of the
potentially infinite quantity $H[\Future]$.

To emphasize the point, it is incorrect to continue the same strategy, however.
That is, picking up from Eq.~(\ref{eq:FStateFuture}) the following is
ambiguous:
\begin{align*}
I[\Past;\Future] & = I[\FutureCausalState;\Future] \\
  & = I[\FutureCausalState;\PastEps(\Future)] \\
  & = I[\FutureCausalState;\PastCausalState]
\end{align*}
even though the final line is the goal and, ultimately, is correct. Why? To
see this, we again expand out the intermediary steps implied:
\begin{align}
I[\FutureCausalState;\Future] & = H[\Future] - H[\Future|\FutureCausalState]
  \label{eq:GetToFutureConditionedOnState} \\
  & = H[\PastEps(\Future)] - H[\PastEps(\Future)|\FutureCausalState]
  	\label{eq:WrongExpansion} \\
  & = H[\PastCausalState] - H[\PastCausalState|\FutureCausalState]
  \nonumber \\
  & = I[\PastCausalState;\FutureCausalState] ~.
  \nonumber
\end{align}
That second step (Eq.~(\ref{eq:WrongExpansion})), by violating the rule of
matching objects types, is wrong. And so, the ensuing steps do not follow,
even if the desired result is obtained.

The conclusion is that the second use of the causal equivalence relation,
seemingly forced in the original short proof of Thm.~\ref{EasCausalMI}, is
not valid. The solution is to find a different proof strategy that does not
lead to this cul de sac.

There is an alternative expansion to Eq.
(\ref{eq:GetToFutureConditionedOnState}) that appears to avoid the problem:
\begin{align}
I[\FutureCausalState;\Future]
  & = H[\FutureCausalState] - H[\FutureCausalState|\Future] \nonumber \\
  & = H[\FutureCausalState] - H[\FutureCausalState|\PastEps(\Future)] 
  \label{eq:rulecounterex}\\
  & = H[\FutureCausalState] - H[\FutureCausalState|\PastCausalState] \nonumber \\
  & = I[\PastCausalState;\FutureCausalState] \nonumber ~.
\end{align}
This seems fine, since no overtly infinite quantities appear and $\PastEps(\cdot)$ is used only in conditioning.

The step to Eq.~(\ref{eq:rulecounterex}) is still problematic, though. The
concern is that, on the one hand, the retrodictive causal states are sufficient
for the pasts, as indicated in Eq.~(\ref{eq:Sufficient}). On the other hand,
it does not immediately follow that they are sufficient for predictive causal
states, as required by Eq.~(\ref{eq:rulecounterex}).

In short, these problems result from ignoring that the goal involves a
higher-dimensional, multivariate problem. We need a
strategy that avoids the ambiguities and gives a reliable procedure. This is
found in using the four-variable mutual informations introduced in
Refs.~\cite{Crut08a} and \cite{Crut08b}. This is the strategy we now lay out
and it also serves to illustrate the rules required for the more concise proof
strategy.

\section{Detailed Proof}

In addition to the rule of not introducing incomparable objects, we need
several basic results.
First, the causal equivalence relations lead to the informational identities:
\begin{align*}
H[\FutureCausalState|\Past] & = 0 ~,\\
H[\PastCausalState|\Future] & = 0 ~.
\end{align*}
That is, these state uncertainties vanish, since $\FutureEps(\cdot)$ and
$\PastEps(\cdot)$ are functions, respectively, of the past and future.

Second, causal states have the Markovian property that they render the past and
future statistically independent. They \emph{causally shield} the future from
the past:
\begin{align*}
I[\Past;\Future|\FutureCausalState] & = 0 ~,\\
I[\Past;\Future|\PastCausalState] & = 0 ~.
\end{align*}
In this way, one sees how the causal states are the structural decomposition
of a process into conditionally independent modules. Moreover, they are defined
to be optimally predictive in the sense that knowing which causal state a 
process is in is just as good as having the entire past in hand:
$\Prob(\Future|\FutureCausalState) = \Prob(\Future|\Past)$ or,
equivalently, $\EE = I[\FutureCausalState;\Future]$.

Now, we consider several additional identities that follow more or less
straightforwardly from the \eM's defining properties.
\begin{Lem}
$I[\CSjoint[;]|\Past] = 0$
and $I[\CSjoint[;]|\Future] = 0$.
\label{lem:SpastSfutGivenPast}
\end{Lem}

\begin{ProLem}
These vanish since the past (future) determines the predictive (retrodictive)
causal state.
\qed
\end{ProLem}

\begin{Lem}
$I[\Past;\Future;\PastCausalState|\FutureCausalState] = 0$.
\label{lem:PastSPastFutureGivenSFuture}
\end{Lem}

\begin{ProLem}
\begin{align*}
I[\Past;\Future;\FutureCausalState|\PastCausalState]
  & = I[\Past;\Future|\PastCausalState]
    - I[\Past;\Future|\CSjoint[,]] \\
  & = 0 - 0 ~.
\end{align*}
The terms vanish by causal shielding.
\qed
\end{ProLem}

\begin{Lem}
$I[\CSjoint[;];\Future|\Past] = 0$.
\label{lem:SpastSfutFutureGivenPast}
\end{Lem}

\begin{ProLem}
\begin{align*}
I[\CSjoint[;];\Future|\Past]
  & = I[\CSjoint[;]|\Past] \\
  & ~~~ - I[\CSjoint[;]|\Past,\Future] ~.
\end{align*}
The first term vanishes by Lemma~\ref{lem:SpastSfutGivenPast}.
Expanding the second term we see that:
\begin{align*}
I[\CSjoint[;]|\Past,\Future]
  & = H[\FutureCausalState|\Past,\Future] \\
  & ~~~~ - H[\FutureCausalState|\Past,\Future,\PastCausalState] ~.
\end{align*}
Both terms here vanish since the past determines the predictive causal state.
\qed
\end{ProLem}

Now, we are ready for the proof. First, recall the theorem's statement.

\begin{The}
\label{EasCausalMI}
Excess entropy is the mutual information between the predictive
and retrodictive causal states:
\begin{equation}
\EE = I[\CSjoint[;]] ~.
\end{equation}
\end{The}

\begin{ProThe}
This follows via a parallel reduction of the four-variable mutual information
$I[\Past;\Future;\CSjoint[;]]$ into
$I[\Past;\Future]$ and $I[\FutureCausalState;\PastCausalState]$.
The first reduction is:
\begin{align*}
I[\Past;\Future;\CSjoint[;]] 
  & = I[\Past;\Future;\FutureCausalState]
    - I[\Past;\Future;\FutureCausalState|\PastCausalState] \\
  & = I[\Past;\Future;\FutureCausalState] \\
  & = I[\Past;\Future]
    - I[\Past;\Future|\FutureCausalState] \\
  & = I[\Past;\Future] \\
  & = \EE ~.
\end{align*}
The second line follows from Lemma~\ref{lem:PastSPastFutureGivenSFuture}
and the fourth from causal shielding.

The second reduction is, then:
\begin{align*}
I[\Past;\Future;\CSjoint[;]] 
  & = I[\CSjoint[;];\Future]
    - I[\CSjoint[;];\Future|\Past] \\
  & = I[\CSjoint[;];\Future] \\
  & = I[\CSjoint[;]]
    - I[\CSjoint[;]|\Future] \\
  & = I[\CSjoint[;]] ~.
\end{align*}
The second line follows from Lemma~\ref{lem:SpastSfutFutureGivenPast}
and the fourth from Lemma~\ref{lem:SpastSfutGivenPast}.
\qed
\end{ProThe}

\begin{Rem}
Note that the steps here do not force one into inadvertently using the causal
equivalence relation to introduce incomparable objects.
\end{Rem}

\section{Finite Pasts and Futures}

This is all well and good, but there is a nagging concern in all of the above.
As noted at the beginning, we are improperly using entropies of semi-infinite
chains of random variables. These entropies typically are infinite and so many
of the steps are literally not correct. Fortunately, as we will show, this
concern is so directly addressed that there is rarely an inhibition in the
above uses. The shortcuts that allow their use are extremely handy and allow
much progress and insight, if deployed with care. Ultimately, of course, one
must still go through proofs using proper objects and manipulations and
verifying limits. We now show how to address this issue, highlighting a number
of technicalities that distinguish between important process classes.

The strategy is straightforward, if somewhat tedious and obfuscating: Define
pasts, futures, and causal states over finite-length sequences.  

\begin{Def}
\label{FPCS}
Given a process $\Pr(\BiInfinity)$, its \emph{finite predictive causal states}
$\FutureCausalState_{KL}$ are defined by:
\begin{align*}
\FutureEps_{KL}(\finpast{K}) \equiv \left\{ \finpastp{K} :
  \Prob(\FinFuture{L} | \finpast{K}) = 
  \Prob(\FinFuture{L} | \finpastp{K} ) \right\} ~.
\end{align*}
\end{Def}

\begin{Def}
\label{FRCS}
Given a process $\Pr(\BiInfinity)$, its \emph{finite retrodictive causal states}
$\PastCausalState_{KL}$ are defined by:
\begin{align*}
\PastEps_{KL}(\finfuture{L}) \equiv \left\{ \finfuturep{L} :
  \Prob(\FinPast{K} | \finfuture{L} ) = 
  \Prob(\FinPast{K} | \finfuture{L} ) \right\} ~.
\end{align*}
\end{Def}

That is, we now partition finite pasts (futures) of length $K$ ($L$) with
probabilistically distinct  distributions over finite futures (pasts). We end
up with two sets, $\FutureCausalStateSet_{KL}$ and $\PastCausalStateSet_{KL}$,
which describe the finite-length predictive and retrodictive causal states for
each value of $K$ and $L$.

\begin{Rem}
The subscripts on $\FutureCausalState_{KL}$ and $\PastCausalState_{KL}$ should
not be interpreted as time indices, as they are more commonly used in the
literature.
\end{Rem}

\begin{Rem}
A central issue here is that, in general, for the causal states
$\FutureCausalState$ defined by Eq. (\ref{eq:CausalEquiv}):
\begin{equation}
\FutureCausalState \neq \lim_{K,L\rightarrow\infty}
  \FutureCausalState_{KL} ~.
\end{equation}
The analogous situation is true for $\PastCausalState$. Why? For some processes,
it can happen that $|\FutureCausalStateSet_{KL}| \to \infty$ even though
$|\FutureCausalStateSet| < \infty$.  The result is that the causal states $\FutureCausalState$ are not reached in the above limiting 
procedure. However, their information content can be the same.  And so,
in the following, we must take care in establishing results regarding the
large-$K$ and -$L$ limits.
\end{Rem}

A first example of this is to explain why the applications of
$\epsilon(\cdot)$ in Eqs. (\ref{eq:FirstApplicationOfEps}) and
(\ref{eq:SecondApplicationOfEps}) are plausible.  We establish the
finite-length version of those steps.

\begin{Prop}
$H[\FinFuture{L}|\FinPast{K}] = H[\FinFuture{L}|\FutureCausalState_{KL}]$.
\label{prop:FiniteLengthPastToState}
\end{Prop}

\begin{ProProp}
We calculate directly:
\begin{align*}
H[ & \FinFuture{L} |\FinPast{K}] \\
  & = \sum_{\mathclap{w \in \MeasAlphabet^K}}
  \Prob(w) H[\FinFuture{L}| \FinPast{K} = w] \\
  & = \sum_{\mathclap{w \in \MeasAlphabet^K}}
  \Prob(w)
  H[\FinFuture{L} |
  \FutureCausalState_{KL} = \FutureEps_{KL} (w) ] \\
  & = \sum_{\mathclap{w \in \MeasAlphabet^K}}
  \Prob(w)
  H[\FinFuture{L} |
  \FutureCausalState_{KL} = \FutureEps_{KL} (w) ]
  \sum_{ \mathclap{\causalstate \in \FutureCausalStateSet_{KL}} }
  \delta_{\causalstate, \FutureEps_{KL} (w)}
  \\
  & =
  \sum_{\mathclap{ \causalstate \in \FutureCausalStateSet_{KL} }}
  H[\FinFuture{L} | \FutureCausalState_{KL} = \causalstate ]
  \sum_{\mathclap{ w \in \MeasAlphabet^K }}
  \Prob(w)
  \delta_{\causalstate, \FutureEps_{KL} (w)}
  \\
  & =
  \sum_{\mathclap{ \causalstate \in \FutureCausalStateSet_{KL} }}
  H[\FinFuture{L} | \FutureCausalState_{KL} = \causalstate ]
  \Prob(\causalstate)
  \\
  & =
  H[\FinFuture{L} | \FutureCausalState_{KL} ] ~.
  \hspace{1.5in} \qed
\end{align*}
\end{ProProp}

And so, for all $L$, we have:
\begin{align*}
 H[\Future^L| \FutureCausalState_{\infty L}] 
  &\equiv \lim_{K\rightarrow\infty} H[\Future^L|\FutureCausalState_{K L}]  \\
  &= \lim_{K\rightarrow\infty} H[\Future^L|\Past^K] \\
  &= H[\Future^L|\Past] ~.
\end{align*}
The last step requires a measure-theoretic justification. This is given using
the method of Ref.~\cite[Appendix]{Trav10d}.

\begin{Cor}
$I[\FinPast{K};\FinFuture{L}] = I[\FutureCausalState_{KL};\FinFuture{L}]$.
\end{Cor}

\begin{ProCor}
Following a finite-lengths version of Eq. (\ref{eq:FirstApplicationOfEps}),
we apply Prop. \ref{prop:FiniteLengthPastToState}.
\end{ProCor}

By similar reasoning in the proposition and corollary we have the
time-reversed analogs:
\begin{align*}
H[\FinPast{K}|\FinFuture{L}] & = H[\FinPast{K}|\PastCausalState_{KL}] ~,\\
I[\FinPast{K};\FinFuture{L}] & = I[\FinPast{K};\PastCausalState_{KL}] ~.
\end{align*}

\begin{Def}
The \emph{finite-length excess entropy} is:
\begin{align*}
\EE(K,L) \equiv I[\FinPast{K};\FinFuture{L}] ~.
\end{align*}
\end{Def}

\begin{Lem}
$\EE = \lim_{ K,L \to \infty } \EE(K,L)$.
\label{EasLimPastKFutureL}
\end{Lem}

\begin{ProLem}
It is known that $I[\Past^L;\Future^L]$ converges to $\EE$
\cite{Crut01a,Trav10b}. Thus, it follows straightforwardly that
$I[\Past^K; \Future^L]$ also converges to $\EE$, so long as $K$ and $L$
simultaneously diverge to infinity.
\qed
\end{ProLem}

We are now, finally, ready to focus in more directly on the original goal.

\begin{Prop}
$\EE(K,L) = I[\CSjointKL[;]{KL}{KL}]$.
\label{FiniteEasCausalMI}
\end{Prop}

\begin{ProProp}
The proof relies on finite-length analogs to
Lemmas~\ref{lem:SpastSfutGivenPast}, 
\ref{lem:PastSPastFutureGivenSFuture}, and \ref{lem:SpastSfutFutureGivenPast} 
and then proceeds similarly to Thm.~\ref{EasCausalMI}. Specifically,
\begin{align*}
I[\FinPast{K};\CSjointKL[;]{KL}{KL};\FinFuture{L}] 
  & = I[\FinPast{K};\FinFuture{L}]
\intertext{follows from the first reduction in the proof of Thm.~\ref{EasCausalMI} and:}
I[\FinPast{K};\CSjointKL[;]{KL}{KL};\FinFuture{L}] 
  & = I[\CSjointKL[;]{KL}{KL}]
\end{align*}
follows from the second reduction there. All that is changed in the reductions
is the substitution of finite-length quantities. Otherwise, the
information-theoretic identities hold as given there.
\qed
\end{ProProp}

\begin{The}
The excess entropy is:
\begin{align*}
\EE = \lim_{ K,L \to \infty }
  I[\CSjointKL[;]{KL}{KL}] ~.
\end{align*}
\label{thm:EEAsCSLimit}
\end{The}

\begin{ProThe}
By Lemma~\ref{EasLimPastKFutureL}, we relate $\EE$ to the sequence of mutual
informations between the finite past and finite future. By 
Prop.~\ref{FiniteEasCausalMI}, this limit is also equal to the limit of mutual
informations between the finite predictive and finite retrodictive causal
states.
\qed
\end{ProThe}

\begin{Rem}
As with Lemma~\ref{EasLimPastKFutureL}, the limits in $K$ and $L$ must be 
done simultaneously.
\end{Rem}

At this point, we have gone as far as possible, it seems, in relating the
finite-length excess entropy and forward-reverse causal-state mutual
informations. From here on, different kinds of process have different limiting
behaviors. We discuss one such class and so establish the original claim.

Recall the class of processes that can be represented by
\emph{exactly synchronizing \eMs}. Roughly speaking, such a process has an
\eM\ to which an observer comes to know its internal state from a finite
number of measurements. (For background see Ref.~\cite{Trav10b}.)
This is the class of processes we focus on in the following.

\begin{Lem}
If $\FutureEM$ and $\PastEM$ are both exactly synchronizing and each has a 
finite number of (recurrent) causal states, then:
\begin{align}
  I[\CSjoint[;]] = \lim_{K,L\rightarrow\infty} I[\CSjointKL[;]{KL}{KL}] ~.
\label{CSeqCSKL}
\end{align}
\label{thm:ExactSyncProcCSs}
\end{Lem}

\begin{ProLem}
Finitary processes that are exactly synchronizable have at least one
finite-length synchronizing word. And this sync word occurs in almost every
sufficiently long sequence. Thus, as $K$ and $L$ simultaneously tend to
infinity, one eventually constructs a partition that includes a synchronizing
word. From there on, increasing $K$ and $L$ eventually discovers all 
infinite-length causal states, which are finite in number by assumption.  
The result is that probability accumulates in the subset of finite-length 
causal states which correspond to the causal states which are
reachable, infinitely-preceded, and recurrent~\cite{Uppe97a}.  Thus, 
the limit of the finite-length causal states differs from the infinite-length 
causal states only on a set of measure zero. Finally, also by assumption,
this holds for both the forward and reverse \eMs. And so, the information
content in the finite-length causal states limits on the information content 
of the causal states which, by Eq.~(\ref{eq:CausalEquiv}),
are defined in terms of semi-infinite pasts and futures.
\qed
\end{ProLem}

\begin{The}
If $\FutureEM$ and $\PastEM$ are both exactly synchronizing and each has a
finite number of (recurrent) causal states, then:
\begin{align*}
  \EE = I[\CSjoint[;]] ~.
\end{align*}
\label{Thm:ExactSyncEasMutualI}
\end{The}

\begin{ProThe}
Directly from Thm. \ref{thm:EEAsCSLimit} and Lemma \ref{thm:ExactSyncProcCSs}.
\end{ProThe}

\section{Conclusion}

In the preceding, we examined an evocative and, in its simplicity,
innocent-looking identity: $\EE = I[\CSjoint[;]]$.
It tells us that the excess entropy is equal to the mutual information 
between the predictive and retrodictive causal states.
It begins to reveal its subtleties when one realizes that excess entropy is 
defined \emph{solely} in terms of the observed process $\Pr(\Past; \Future)$
and makes no explicit reference to the process's internal organization. 
Additionally, $\Past$ and $\Future$ are continuous
random variables, when $\FutureCausalState$ and $\PastCausalState$
need not be.

In explicating their relationships, finite-length counterparts to the predictive
and retrodictive causal states were introduced, and the limit was taken as the
finite-lengths tended to infinity. A priori, there is no reason to expect that
the finite-length causal states will limit on the causal states, since the
latter are defined over infinite histories and futures. In fact, there are finitary
processes for which the number of finite-length causal states diverges, even
when the number of (asymptotic, recurrent) causal states is finite.

However, when considering exactly synchronizing \eMs, there exists a subset of
the finite-length causal states at each $K$ and $L$ that \emph{does} limit on
the causal states.  When such \eMs\ have a finite number of causal states, it
is possible to identify this subset.  This fact was used to prove
Thm.~\ref{thm:ExactSyncProcCSs}.

When this subset of the finite-length causal states cannot be identified or
when it does not exist, it is still expected that the limit of mutual
informations between the finite-length causal states will equal 
the mutual information between the predictive and retrodictive causal states.
However, the proof for this requires more sophistication and the technique for 
calculating $\EE$, outlined in Ref.~\cite{Crut08b}, needs refining. The set of
\eMs\ that are not exactly synchronizing are among those that would benefit
from such analysis.

\begin{figure}[th]
\begin{center}
\resizebox{!}{1.6in}{\includegraphics{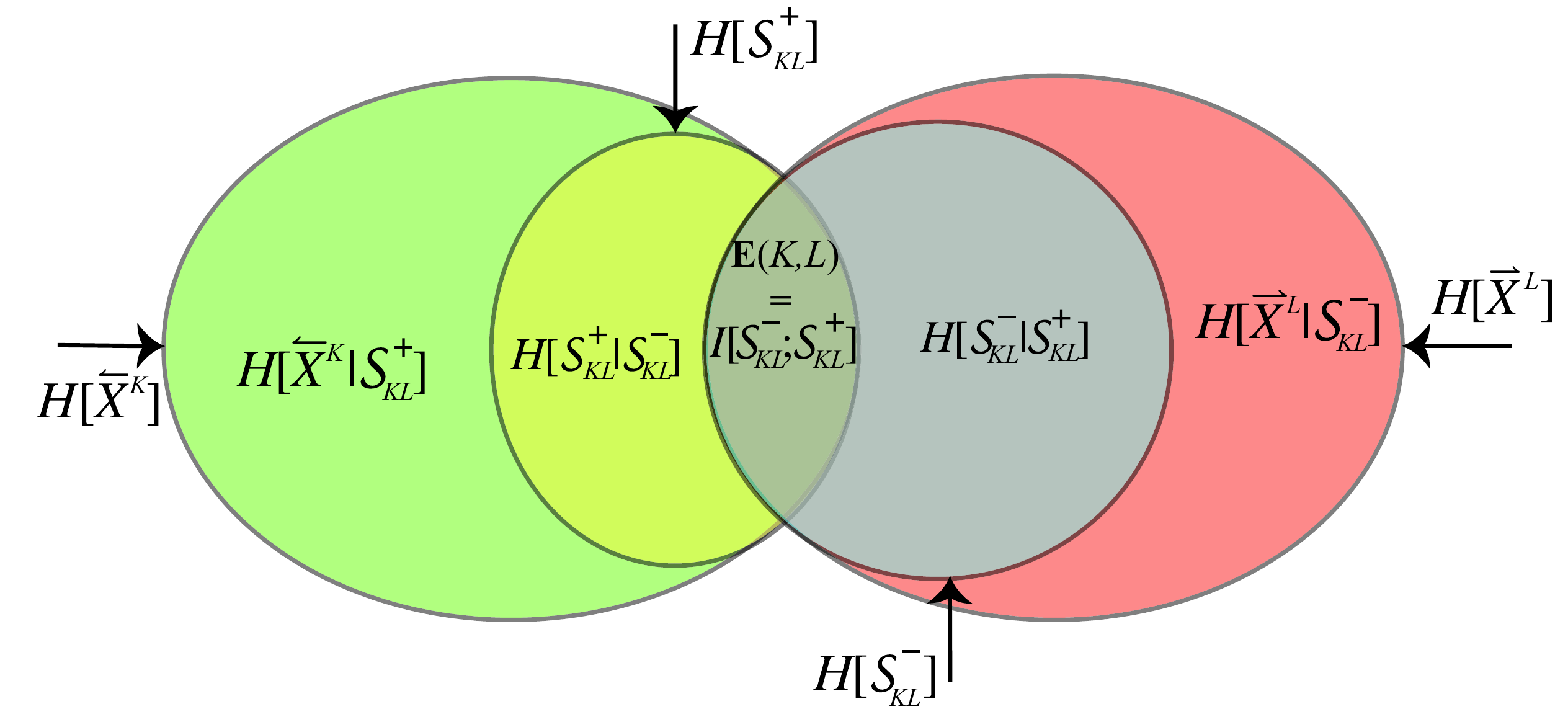}}
\end{center}
\caption{
  \EM\ information diagram over finite length-$K$ past and length-$L$ future
  sequences for a stationary stochastic process.
  }
\label{fig:eMIDiagram}
\end{figure}

The information diagram of Figure~\ref{fig:eMIDiagram} closes our development
by summarizing the more
detailed finite-history and -future framework introduced here. The various
lemmas, showing that this or that mutual information vanished, translate into
information-measure atoms having zero area. The overall diagram is quite similar
to that introduced in Ref.~\cite{Crut08a}, which serves to emphasize the point
made earlier that working with infinite sequences preserves many of the
central relationships in a process's information measure. It also does not
suffer from the criticism, as did the previous one, of representing infinite
atoms as finite.

The information diagram graphically demonstrates that, as done in the detailed
proof given for Thm. \ref{Thm:ExactSyncEasMutualI}, one should avoid using
potentially infinite quantities, such as $H[\Past]$ and $H[\Future]$, whenever
possible, in favor of alternative finite atoms, which are various mutual
informations and conditional mutual informations. Moreover, when infinite atoms
cannot be avoided, then the finite-length quantities must be used and their limits
carefully taken, as we showed.

\section*{Acknowledgments}

We thank Susanne Still for insisting on our being explicit. This work was
partially supported by the Defense Advanced Research Projects Agency
(DARPA) Physical Intelligence project via subcontract No.~9060-000709. The
views, opinions, and findings contained here are those of the authors and
should not be interpreted as representing the official views or policies, either
expressed or implied, of the DARPA or the Department of Defense.

\bibliography{chaos}

\begin{thebibliography}{10}

\bibitem{Crut08a}
J.~P. Crutchfield, C.~J. Ellison, and J.~R. Mahoney.
\newblock Time's barbed arrow: {Irreversibility}, crypticity, and stored
  information.
\newblock {\em Phys. Rev. Lett.}, 103(9):094101, 2009.

\bibitem{Fras86a}
A.~Fraser and H.~L. Swinney.
\newblock Independent coordinates for strange attractors from mutual
  information.
\newblock {\em Phys. Rev. A}, 33:1134--1140, 1986.

\bibitem{Casd91a}
M.~Casdagli and S.~Eubank, editors.
\newblock {\em Nonlinear Modeling}, SFI Studies in the Sciences of Complexity,
  Reading, Massachusetts, 1992. Addison-Wesley.

\bibitem{Spro03a}
J.~C. Sprott.
\newblock {\em Chaos and Time-Series Analysis}.
\newblock Oxford University Press, Oxford, UK, second edition, 2003.

\bibitem{Kant06a}
H.~Kantz and T.~Schreiber.
\newblock {\em Nonlinear Time Series Analysis}.
\newblock Cambridge University Press, Cambridge, UK, second edition, 2006.

\bibitem{Crut97a}
J.~P. Crutchfield and D.~P. Feldman.
\newblock Statistical complexity of simple one-dimensional spin systems.
\newblock {\em Phys. Rev. E}, 55(2):R1239--R1243, 1997.

\bibitem{Erb04a}
I.~Erb and N.~Ay.
\newblock Multi-information in the thermodynamic limit.
\newblock {\em J. Stat. Phys.}, 115:949--967, 2004.

\bibitem{Tono94a}
G.~Tononi, O.~Sporns, and G.~M. Edelman.
\newblock A measure for brain complexity: {Relating} functional segregation and
  integration in the nervous system.
\newblock {\em Proc. Nat. Acad. Sci. USA}, 91:5033--5037, 1994.

\bibitem{Bial00a}
W.~Bialek, I.~Nemenman, and N.~Tishby.
\newblock Predictability, complexity, and learning.
\newblock {\em Neural Computation}, 13:2409--2463, 2001.

\bibitem{Ebel94c}
W.~Ebeling and T.~Poschel.
\newblock Entropy and long-range correlations in literary english.
\newblock {\em Europhys. Lett.}, 26:241--246, 1994.

\bibitem{Debo08a}
L.~Debowski.
\newblock On the vocabulary of grammar-based codes and the logical consistency
  of texts.
\newblock 2008.
\newblock submitted; arXiv.org:0810.3125 [cs.IT].

\bibitem{Crut08b}
C.~J. Ellison, J.~R. Mahoney, and J.~P. Crutchfield.
\newblock Prediction, retrodiction, and the amount of information stored in the
  present.
\newblock {\em J. Stat. Phys.}, 136(6):1005--1034, 2009.

\bibitem{Cove06a}
T.~M. Cover and J.~A. Thomas.
\newblock {\em Elements of Information Theory}.
\newblock Wiley-Interscience, New York, second edition, 2006.

\bibitem{Yeun91a}
R.~Yeung.
\newblock A new outlook on {Shannon}'s information measures.
\newblock {\em IEEE Trans. Info. Th.}, 37(3):466--474, 1991.

\bibitem{Crut01a}
J.~P. Crutchfield and D.~P. Feldman.
\newblock Regularities unseen, randomness observed: Levels of entropy
  convergence.
\newblock {\em CHAOS}, 13(1):25--54, 2003.

\bibitem{CompMechMerge}
J.~P. Crutchfield and K.~Young.
\newblock Inferring statistical complexity.
\newblock {\em Phys. Rev. Let.}, 63:105--108, 1989; J. P. Crutchfield,
  \emph{Physica D} {\bf 75} 11--54, 1994; J. P. Crutchfield and C. R. Shalizi,
  \emph{Phys. Rev. E} {\bf 59}(1) 275--283, 1999.

\bibitem{Trav10d}
N.~Travers and J.~P. Crutchfield.
\newblock Equivalence of history and generator epsilon-machines.
\newblock 2010.
\newblock SFI Working Paper 10-12-XXX; arxiv.org:1012.XXXX [XXXX].

\bibitem{Trav10b}
N.~Travers and J.~P. Crutchfield.
\newblock Asymptotic synchronization for finite-state sources.
\newblock {\em submitted}, 2010.
\newblock arxiv:1011.1581 [nlin.CD].

\bibitem{Uppe97a}
D.~R. Upper.
\newblock {\em Theory and Algorithms for Hidden {M}arkov Models and Generalized
  Hidden {M}arkov Models}.
\newblock PhD thesis, University of California, Berkeley, 1997.
\newblock {P}ublished by University Microfilms Intl, Ann Arbor, Michigan.

\end{thebibliography}

\end{document}